\documentclass[aps,prl,amsmath,twocolumn,superscriptaddress,letterpaper,floatfix]{revtex4-1}
\usepackage{graphicx,color}
\usepackage{verbatim}
\usepackage{amssymb}   
\usepackage{amsmath}
\usepackage{amsfonts}
\usepackage{mathrsfs}
\usepackage{mathdots}
\usepackage{hyperref}
\hypersetup{hidelinks}
\usepackage{epsfig}

\usepackage{braket}
\usepackage{bm}
\begin{document}

\title{Probing Antialtermagnetism via Orbital-Field-Induced Spin Splitting}

\author{Zi-Ting Sun}\thanks{Contact author: zsunaw@connect.ust.hk}
 \affiliation{RIKEN Center for Emergent Matter Science (CEMS), Wako, Saitama 351-0198, Japan}

	\date{\today}
\begin{abstract}

Layer compensation can conceal spin polarization behind a spin-degenerate bulk spectrum in antialtermagnets, preventing spectroscopic identification of the underlying magnetic order. Here we show that the orbital effect of an in-plane magnetic field converts the hidden spin texture into an observable spin splitting with momentum parity opposite to that of the underlying exchange order, thereby restoring spectroscopic access. Once the parity is determined, complementary response functions in the weak-field regime further resolve its wave character. Within a minimal model, we illustrate that a sum rule over the difference of spectral functions between opposite spins isolates the momentum gradient of the form factor and differentiates hidden even-parity orders, whereas for odd-parity sectors the induced net spin polarization exhibits characteristic field-amplitude and angular dependences. These results establish orbital-field-induced spin splitting as a generic route to identifying the hidden altermagnetic order parameter without layer resolution.

\end{abstract}
	\maketitle

\emph{Introduction.}---Altermagnetism has attracted intense interest as a third form of collinear magnetism beyond ferromagnetism and
antiferromagnetism \cite{vsmejkal2022emerging,mazin2022altermagnetism}.
Altermagnets combine hallmarks of both: their
electronic bands carry an anisotropic nonrelativistic spin splitting while the
net spin magnetization vanishes
\cite{shim2025spinpolarized,xie2026review}.
Recent developments in spin space group theory
\cite{liu2026oriented,song2026unified,luo2026spin} have extended the research scope of altermagnetism to include noncollinear spin textures and odd-parity spin splitting \cite{hellenes2023exchange,brekke2024minimal,cheong2024noncollinear,yu2025oddparity}.
This enlarged family of unconventional magnetism opens opportunities for
applications such as magnetic memory
\cite{vsmejkal2022giant,song2025functional,song2025electrical}, spintronic
devices
\cite{chakraborty2025highly,fang2024quantum,ouyang2026fingerprints,jungwirth2026spintronics}, and topological quantum
computation
\cite{zhu2023topological,ghorashi2024altermagnetic,sun2025pseudo,yang2025topological,luo2026hidden}.

Observing the anisotropically spin-split spectrum is the most direct route
to identifying an altermagnetic order parameter
\cite{jungwirth2026signatures}, as demonstrated by photoemission in MnTe
\cite{krempasky2024altermagnetic} and by quantum oscillations in
CrSb \cite{long2026bulk}. Such a readout fails,
however, when the spin polarization is hidden
\cite{zhang2014hidden,riley2014direct}. This is the case in an
antialtermagnet (AAM) \cite{meier2026antialtermagnetism}, where the
altermagnetic order parameter of one layer is fully compensated by that of its
partner layer \cite{matsuda2025multiferroic,huang2026hiddenvisible}, realizing a nonrelativistic analogue of the
hidden spin texture in centrosymmetric crystals
\cite{yuan2023uncovering,guo2026hidden,xiong2026matter}. Such materials exhibit spin-degenerate bulk bands throughout the Brillouin zone and are therefore invisible to conventional bulk-sensitive spin spectroscopy \cite{yang2025observation}. Surface-sensitive spin probes can access local spin textures \cite{jiang2025metallic,zhang2025spinvalley}, but complex surface effects can make their connection to the hidden bulk order ambiguous \cite{sun2025kv2se2o,lange2026emergent,xie2026rbv2te2o}. How to reveal the parity and wave character of such a hidden bulk exchange order in an AAM without resorting to local resolution remains an open challenge.

\begin{figure}[t]
		\centering
		\includegraphics[width=1\linewidth]{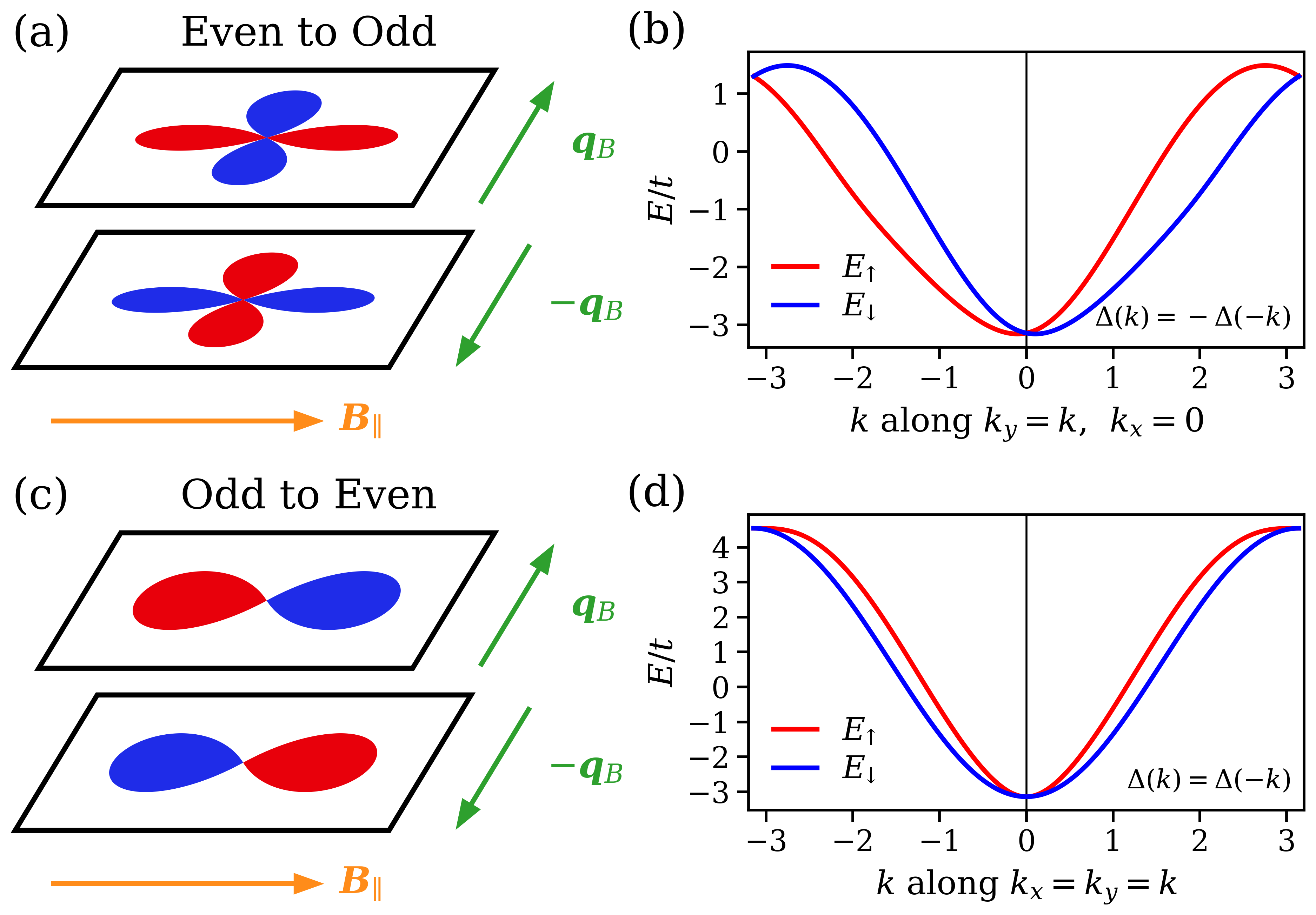}
		\caption{Schematics of orbital-field-induced even-odd parity
conversion in a bilayer antialtermagnet: (a), (b) Even-to-odd conversion: opposite even-parity $d$-wave textures in center-related layers acquire transverse Peierls shifts $\pm\boldsymbol q_B\perp\boldsymbol B_\parallel$, producing an odd-in-momentum spin splitting $\Delta(k)=-\Delta(-k)$.
(c), (d) Odd-to-even conversion: opposite odd-parity $p$-wave textures produce an even-in-momentum splitting $\Delta(k)=\Delta(-k)$.
Red and blue denote spin $\uparrow$ and $\downarrow$ bands, respectively.
Spectra in (b) and (d) are the upper branch ($\nu=+1$) of Eq.~\eqref{eq:bilayer_exact_spectrum} with $t=1$, $t_\perp=0.5$, $J=0.6$, $\mu=-0.2$, $d=1$, $e/\hbar=1$, and $\boldsymbol q_B=(0,-0.4)$.}
		\label{fig:fig1}
\end{figure}

In this Letter, we show that the orbital coupling of an in-plane magnetic
field supplies the missing readout, because it naturally acts as a layer-selective perturbation \cite{xie2023orbital}. For a quick grasp of the physical picture, we consider
a bilayer setting of an AAM shown in Fig.~\ref{fig:fig1}. The field
$\boldsymbol B_\parallel$ imposes opposite transverse Peierls shifts $\pm\boldsymbol q_B\perp\boldsymbol B_\parallel$ on the top and
bottom layers, respectively. The shifts lift the compensation between their
order parameters and produce a field-odd spin splitting
$\Delta(\boldsymbol k;B)\equiv E_{\uparrow}(\boldsymbol k;B)-
E_{\downarrow}(\boldsymbol k;B)$. For a hidden exchange order of definite
parity $\chi=\pm1$, as we prove below, the induced splitting obeys
$\Delta(-\boldsymbol k;B)=-\chi\Delta(\boldsymbol k;B)$ exactly, so that an
even-parity order is converted to an odd splitting and vice versa
[Figs.~\ref{fig:fig1}(a)--(d)]. In this way,
the in-plane orbital field maps the identification of a hidden
nonrelativistic spin texture back onto a measurement of a spin-split
spectrum. Once the parity has been established, the wave character can be further resolved by studying how the induced splitting depends on the
magnitude and the direction of the field. These features allow us to use the
in-plane orbital field as a generic probe of the hidden order parameter
characteristic of an AAM.

\emph{Antialtermagnet with spin degeneracy.}---We first formulate the
generic model for a spin-degenerate AAM. Consider an AAM thin film with an
even number of layers $L=2N$. We label the layers by
$l=1,\ldots,2N$ along $z$ and denote the center-related partner of layer $l$
by $\bar l=2N+1-l$. For a uniformly spaced stack, the coordinate
of layer $l$, measured from the film midplane, is
$z_l=d[l-(2N+1)/2]$, where $d$ is the interlayer spacing; hence,
$z_{\bar l}=-z_l$. We set the in-plane lattice constant $a$ to unity. We further assume that spin-orbit coupling is much weaker than the nonrelativistic exchange order, so that spin $s=\pm1$ along the collinear axis can be taken as conserved. Let $c_{\boldsymbol kls}$ collect all the orbital and sublattice annihilation
operators in layer $l$ at fixed spin $s$, and define
$\Psi_{\boldsymbol ks}=(c_{\boldsymbol k1s}^T,\ldots,
c_{\boldsymbol k,2N,s}^T)^T$. Thus the second-quantized Hamiltonian of the AAM is
$\mathcal H_0=\sum_{\boldsymbol k,s}\Psi_{\boldsymbol ks}^{\dagger}
h_s^{(0)}(\boldsymbol k)\Psi_{\boldsymbol ks}$, in which
\begin{equation}
h_s^{(0)}(\boldsymbol k)=h_0(\boldsymbol k)+sM(\boldsymbol k),
\label{eq:general_2N_parent}
\end{equation}
where $h_0$ is the spin-independent part and $M$ is the exchange part of the Bloch Hamiltonian $h_s^{(0)}(\boldsymbol k)$. Denoting $\mathcal L_l$ as the set of Wannier orbitals $|w_a\rangle$ centered in
layer $l$, and defining the layer projector \cite{hu2026theory}
$P_l=\sum_{a\in\mathcal L_l}|w_a\rangle\langle w_a|$, which obeys
$P_lP_{l'}=\delta_{ll'}P_l$ and $\sum_lP_l=\mathbb I$, we can resolve $h_s^{(0)}$ into the layer blocks as $h_s^{(0)}=\sum_{ll'}h_{s,ll'}^{(0)}$ with
$h_{s,ll'}^{(0)}\equiv P_lh_s^{(0)}P_{l'}=[h_0]_{ll'}+s[M]_{ll'}$.

The spin degeneracy of the model in Eq.~\eqref{eq:general_2N_parent} follows from a center-exchange operation
$\mathcal R_z$ that maps each layer onto its partner. This operation exchanges only the layer projectors
$P_l\leftrightarrow P_{\bar l}$ and satisfies
$\mathcal R_z h_0(\boldsymbol k)\mathcal R_z^{-1}=h_0(\boldsymbol k)$ and $\mathcal R_zM(\boldsymbol k)\mathcal R_z^{-1}=-M(\boldsymbol k)$. In particular, the second condition encodes the key characteristic of an AAM: the altermagnetic exchange order is stacked in a layer-staggered way, and $\mathcal R_z$ enforces compensation between the two partner layers, with $\mathrm{Tr}M(\boldsymbol k)=0$. Here and below, the trace is over the layer, orbital, and
sublattice space of one spin block. Together, these conditions map $h_s^{(0)}(\boldsymbol k)$ to $h_{-s}^{(0)}(\boldsymbol k)$, rendering every band spin degenerate. We classify the parity of the AAM order parameter directly by the momentum parity of its exchange matrix \cite{luo2026spin}, $M(-\boldsymbol k)=\chi M(\boldsymbol k)$: $\chi=+1$ defines an even-parity
AAM, whereas $\chi=-1$ defines an odd-parity AAM.

\emph{Orbital-field-induced spin splitting.}---An in-plane orbital field removes this spectral cancellation. For
$\boldsymbol B_\parallel$, choose the vector potential as $\boldsymbol A_l=z_l\boldsymbol B_\parallel\times\hat{\boldsymbol z}$. A block
connecting layers $l,l'$ then acquires
the Peierls momentum shift
$\boldsymbol q_{ll'}=e\boldsymbol B_\parallel\times\hat{\boldsymbol z}(z_l+z_{l'})/(2\hbar)$, giving \cite{hu2026theory}
\begin{equation}
 [h_s^B(\boldsymbol k)]_{ll'}=
 [h_0(\boldsymbol k+\boldsymbol q_{ll'})
 +sM(\boldsymbol k+\boldsymbol q_{ll'})]_{ll'}.
 \label{eq:AAM_field_dressed_H}
\end{equation}
Because $z_{\bar l}=-z_l$, one has
$\boldsymbol q_{\bar l\bar l'}=-\boldsymbol q_{ll'}$. Together with the
zero-field $\mathcal R_z$ conditions, this yields
$\mathcal R_z h_s^B(\boldsymbol k)\mathcal R_z^{-1}
=h_{-s}^{-B}(\boldsymbol k)$. Here $B$ denotes the signed field amplitude
along a fixed in-plane direction. Applying the operator identity to an
eigenstate gives $E_{m,s}(\boldsymbol k;B)=E_{m,-s}(\boldsymbol k;-B)$, where
$m$ is a band label. Hence the signed splitting
$\Delta_m=E_{m,+}-E_{m,-}$ satisfies
$\Delta_m(\boldsymbol k;-B)=-\Delta_m(\boldsymbol k;B)$. Thus the splitting is
odd under field reversal. At a fixed nonzero $B$, however, $\mathcal R_z$ symmetry is broken, and a
finite spin splitting is generically allowed.

\emph{Symmetry-enforced parity reversal.}---The central result requires one additional ingredient: a symmetry that fixes the parity of the hidden order. At zero field, the AAM has three classes of coexisting spin space group
symmetries, denoted $\mathcal A$, $\mathcal B$, and $\mathcal C$, which establish spectral correspondences between $(\boldsymbol k,s)$ and $(\boldsymbol k,-s)$, $(-\boldsymbol k,s)$, $(-\boldsymbol k,-s)$, respectively. Any two imply the third. Here $\mathcal A=[C_2^\perp\Vert\mathcal R_z]$ protects the
spin-degenerate spectrum, while a layer-preserving operation $\mathcal S$
realizes $\mathcal B$ for even parity or $\mathcal C$ for odd parity. Its
operator realizations, including both unitary and antiunitary cases, are given in
\textbf{End Matter App.~A}. In either case, $\mathcal S$ maps
$h_s^{(0)}(\boldsymbol k)$ to $h_{\chi s}^{(0)}(-\boldsymbol k)$. Acting on the field-dressed Hamiltonian, $\mathcal S$ reverses the field and is
therefore broken, while its product with the center exchange
$\mathcal G=\mathcal R_z\mathcal S$ preserves the applied field and obeys
\begin{equation}
\mathcal G h_s^B(\boldsymbol k)\mathcal G^{-1}
=h_{-\chi s}^{B}(-\boldsymbol k).
\label{eq:main_G_fixed_field}
\end{equation}
For antiunitary $\mathcal S$, the left-hand side includes complex conjugation.
In both cases, Eq.~\eqref{eq:main_G_fixed_field} gives
$E_{m,s}(\boldsymbol k;B)=E_{m,-\chi s}(-\boldsymbol k;B)$ and hence
\begin{equation}
\Delta_m(-\boldsymbol k;B)=-\chi\Delta_m(\boldsymbol k;B).
\end{equation}
This relation is exact for every band: the hidden even-parity order parameter therefore produces an odd-parity, compensated spin splitting, whereas a hidden odd-parity order parameter generates an even-parity splitting. This parity conversion is the primary spectroscopic diagnostic of the parent AAM order.

The same symmetry relation carries over directly to the spin-resolved
spectral function
\begin{equation}
A_s(\mathbf{k},\omega;B)
=
-\frac{1}{\pi}
\operatorname{Im}
\operatorname{Tr}
G_s^R(\mathbf{k},\omega;B),
\end{equation}
where
$G_s^R(\mathbf{k},\omega;B)
=
[\omega+i0^+-h_s^B(\mathbf{k})]^{-1}$. From the spectral relation, one immediately obtains
$A_s(-\mathbf{k},\omega;B)
=
A_{-\chi s}(\mathbf{k},\omega;B)$,
and hence
\begin{equation}
A_{\rm sd}(-\mathbf{k},\omega;B)
=
-\chi A_{\rm sd}(\mathbf{k},\omega;B),
\label{eq:main_spectral_function}
\end{equation}
where $A_{\rm sd}\equiv A_+-A_-$ denotes the spin-difference spectral function.
Thus the parity of the hidden order can be read out directly from
whether $A_{\rm sd}$ reverses or retains its sign between opposite
momenta near the split bands, without assigning individual band indices.

For the odd-to-even conversion, however, the induced spin splitting is not necessarily compensated. Compensation is enforced only when a residual field-preserving
symmetry relates opposite spins. If its momentum action $R_g\neq\pm\mathbb I$
enforces $E_{m,s}(\boldsymbol k;B)=E_{m,-s}(R_g\boldsymbol k;B)$, then
$\Delta_m(R_g\boldsymbol k;B)=-\Delta_m(\boldsymbol k;B)$ and the net spin
polarization vanishes. The field directions where $\Delta_m(\boldsymbol k;B)$ is compensated therefore encode the
angular momentum of odd-parity order (see \textbf{End Matter App.~A}).

\emph{Minimal multilayer model.}---To illustrate the parity theorem explicitly, we consider a minimal multilayer model of an AAM. Its nonmagnetic part is
\begin{equation}
[h_0(\boldsymbol k)]_{ll'}=\varepsilon(\boldsymbol k)\delta_{ll'}
+t_\perp(\delta_{l,l'+1}+\delta_{l,l'-1}).
\label{eq:minimal_h0_blocks}
\end{equation}
Here $\varepsilon(-\boldsymbol k)=\varepsilon(\boldsymbol k)$ is the intralayer
dispersion, and $t_\perp$ is the nearest-layer hopping. For the AAM exchange
order parameter, we choose
\begin{equation}
M(\boldsymbol k)=J f_\alpha(\boldsymbol k)\Lambda.
\label{eq:minimal_M_blocks}
\end{equation}
Here $J$ sets the amplitude of the hidden exchange order, and the alternating projector sum is $\Lambda=\sum_l(-1)^{l-1}P_l$. The exact spectrum of this model is
\begin{equation}
E_{n,\nu,s}^{(0)}=\varepsilon+\nu R_n.
\label{eq:minimal_multilayer_spectrum}
\end{equation}
Here, we define $R_n=(J^2f_\alpha^2+4t_\perp^2\cos^2\theta_n)^{1/2}$ and $\theta_n=n\pi/(2N+1)$, with $n=1,\ldots,N$ labeling the standing-wave subbands quantized along the finite layer direction, while $\nu=+1$ ($-1$) denotes the upper (lower)
layer-hybridized branch within each $n$. The spectrum is independent of $s$, explicitly realizing the general spin-degeneracy condition.

The form factor $f_\alpha$ satisfies $f_\alpha(-\boldsymbol k)=\chi_\alpha f_\alpha(\boldsymbol k)$ and thus has the matrix parity defined above. It determines the momentum dependence of the hidden AAM order parameter, and we use
$\alpha\in\{ s,d,p\}$ to label the $s$-, $d$-, and $p$-wave cases. The $s$-wave case $f_{s}=1$ is momentum independent and therefore represents the ordinary
A-type antiferromagnet (AFM) \cite{mak2019probing}. The even form factor
$f_d=\cos k_x-\cos k_y$ describes a hidden $d$-wave order parameter, whereas the
odd form factor $f_p=\sin k_x$ represents a $p$-wave order parameter of an AAM. The odd-parity order is treated here as an effective spin-conserving exchange texture, which can be generated from the orbital order in a collinear model~\cite{zhuang2025odd,lin2026odd}, or from the low-energy band projection in a noncollinear model~\cite{hellenes2023exchange,sun2025pseudo}.

\begin{figure}[t]
    \centering
    \includegraphics[width=\linewidth]{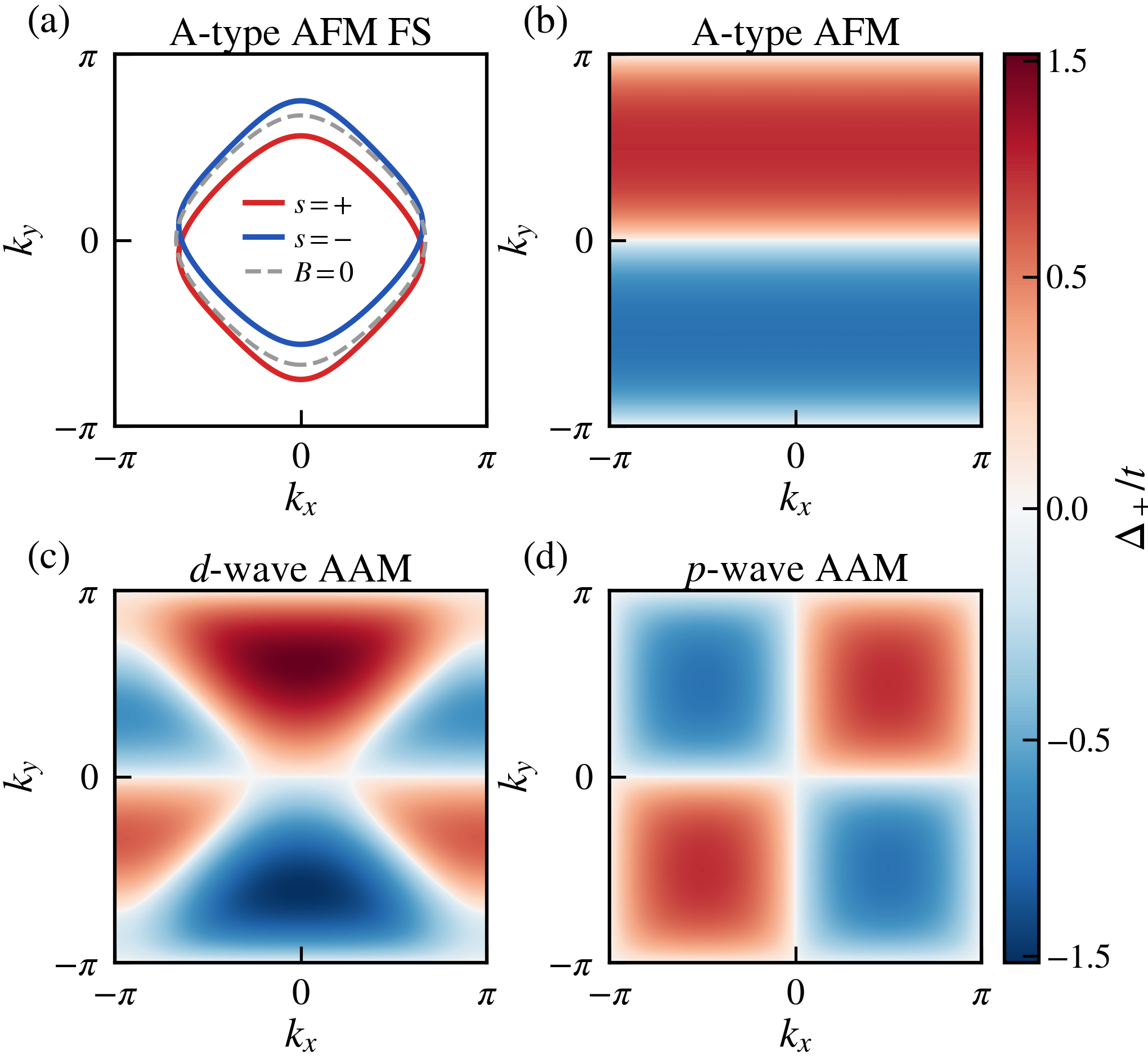}
    \caption{Bilayer illustration of orbital-field-induced spin splitting. (a) Spin-resolved Fermi-surface contours of the $\nu=+1$ branch (red and blue) for the A-type AFM under $\boldsymbol B_\parallel\parallel\hat{\boldsymbol x}$, displaced in opposite directions from the spin-degenerate zero-field contour (gray dashed).
    (b)--(d) Momentum-resolved splitting $\Delta_+(\boldsymbol k)/t$ from Eq.~\eqref{eq:bilayer_exact_splitting}, which encodes the symmetry of the order parameter. The A-type AFM in (b) and $d$-wave AAM in (c) both produce odd-in-momentum splittings, but with distinct wave characters ($p_y$-like and $p+f$-like, respectively); the $p$-wave AAM in (d) produces an even, $d_{xy}$-like splitting. The parameters are identical to those used in Fig.~\ref{fig:fig1}.}
    \label{fig:bilayer_bandmaps}
\end{figure}

\emph{Bilayer illustration.}---We now specialize to the exactly solvable bilayer case, $N=1$. With Pauli matrices $\tau_i$ acting in the layer basis and
$\boldsymbol q_{1 (2)}=\mp\boldsymbol q_B=\mp\boldsymbol B_\parallel\times\hat{\boldsymbol z}\,ed/(2\hbar)$, we define $g_\pm=(g(\boldsymbol k+\boldsymbol q_B)\pm g(\boldsymbol k-\boldsymbol q_B))/2$ for $g=\varepsilon,f_\alpha$. Eq.~\eqref{eq:AAM_field_dressed_H}
then reduces to
\begin{equation}
h_s^B(\boldsymbol k)=
(\varepsilon_{+}-sJf_{\alpha,-})\tau_0+t_\perp\tau_x
+(sJf_{\alpha,+}-\varepsilon_-)\tau_z .
\label{eq:bilayer_field_hamiltonian}
\end{equation}
Defining $R_{\alpha,s}=[t_\perp^2+(\varepsilon_{-}
-sJf_{\alpha,{+}})^2]^{1/2}$, we obtain the exact spectrum
\begin{equation}
E_{\nu,s}^\alpha(\boldsymbol k;B)=
\varepsilon_{+}-sJf_{\alpha,{-}}+\nu R_{\alpha,s},
\qquad \nu=\pm1.
\label{eq:bilayer_exact_spectrum}
\end{equation}
Then the induced spin splitting $\Delta_\nu^\alpha=E_{\nu,+}^\alpha-E_{\nu,-}^\alpha$ reads
\begin{equation}\label{eq:bilayer_exact_splitting}
\Delta_\nu^\alpha = -2 J \left( f_{\alpha,-} + \frac{2 \nu \varepsilon_- f_{\alpha,+}}{R_{\alpha,+} + R_{\alpha,-}} \right).
\end{equation}
The parities of $\varepsilon_{{\pm}}$ and $f_{\alpha,{\pm}}$ imply $\Delta_\nu^\alpha(-\boldsymbol k;B)=-\chi_\alpha
\Delta_\nu^\alpha(\boldsymbol k;B)$, explicitly realizing the parity theorem.

For the $s$-wave case, the relevant $\mathcal B$ symmetry is $[E\Vert C_{2z}]$, which sends $\boldsymbol k\to-\boldsymbol k$ while preserving spin and layer. Taking $\varepsilon=-2t(\cos k_x+\cos k_y)-\mu$ and $\boldsymbol B_\parallel=B\hat{\boldsymbol x}$, i.e., $\boldsymbol q_B=-q_B\hat{\boldsymbol y}$ with $q_B= edB/(2\hbar)$, the induced spin splitting shows the character of a $p_y$-wave magnet, as shown in Figs.~\ref{fig:bilayer_bandmaps}(a) and (b). This can be understood from a small-field expansion of Eq.~\eqref{eq:bilayer_exact_splitting}:
\begin{equation}
\Delta_{\nu}^{s}
\approx
\frac{4\nu Jtq_B}{R_s}\sin k_y
+O(q_B^3),
\label{eq:s_result}
\end{equation}
where $R_\alpha=(t_\perp^2+J^2f_\alpha^2)^{1/2}$. A static external field, rather than optical driving \cite{huang2026lightinduced,zhu2026floquet,liu2026lightinduced}, therefore converts an ordinary A-type AFM into an odd-parity magnet. By contrast, the same field does not lift the spin degeneracy of an ordinary G-type AFM, whose exchange order is compensated not only between neighboring layers but also between opposite Néel sublattices within each layer \cite{mak2019probing}.

The even form factor $f_d$ follows the same even-to-odd route,
as shown in Figs.~\ref{fig:fig1}(a) and (b): A canonical symmetry realization is the planar $d$-wave spin group
${}^{2}4/{}^{1}m{}^{1}m{}^{2}m$ \cite{fang2024quantum}. Its characteristic operation
$[C_2^\perp\Vert C_{4z}]$ exchanges the two spin sectors under a
fourfold rotation and thereby characterizes the $d$-wave spin group,
whereas the in-plane inversion symmetry $[E\Vert C_{2z}]$ supplies the required $\mathcal B$. To linear order in $q_B$, the band splitting is
\begin{equation}
\Delta_{\nu}^{d}
\approx
2Jq_B\sin k_y
\left(
1+\frac{2\nu t f_d}{R_d}
\right).
\label{eq:d_result}
\end{equation}
Near the $\Gamma$ point, the splitting contains an odd $p_y$-wave contribution modified by an additional $f$-wave term. The resulting $p+f$ mixture accounts for the nodal structure in Fig.~\ref{fig:bilayer_bandmaps}(c). Although its splitting is distinct from that of the A-type AFM, both are odd in momentum and therefore cannot be distinguished by parity alone.

For the odd form factor $f_p=\sin k_x$, a characteristic spin-group
operation of the collinear $p_x$-wave odd-parity altermagnet is the
opposite-spin mirror $[C_2^\perp\Vert m_x]$ \cite{zeng2026odd}. Combined with
the same-spin mirror $[E\Vert m_y]$, it gives
$[C_2^\perp\Vert C_{2z}]$, thereby realizing the full
momentum-reversing correspondence $\mathcal C$ required by the
odd-to-even route. For $\boldsymbol B_\parallel\parallel\hat{\boldsymbol x}$,
$[C_2^\perp\Vert m_x]$ itself preserves the axial field and enforces
$E_{\nu,s}(k_x,k_y;B_x)
=
E_{\nu,-s}(-k_x,k_y;B_x)$,
so that the induced spin splitting remains
compensated, as shown in Figs.~\ref{fig:fig1}(c) and (d). Figure~\ref{fig:bilayer_bandmaps}(d) shows the
resulting even-parity splitting with a $d_{xy}$-like pattern. To linear order in $q_B$, the splitting is
\begin{equation}
\Delta_{\nu}^{p}
\approx
\frac{4\nu Jtq_B}{R_p}\sin k_x\sin k_y.
\label{eq:p_result}
\end{equation}
A generic field
direction breaks this residual protection and therefore permits a net spin polarization.

\emph{Sum rule diagnostic of hidden exchange order.}---The exact parity reversal
provides the first diagnostic. To distinguish orders within the same parity
sector, we examine the weak-field spin splitting more closely.

The weak-field response is set by an orbital moment built from two operators: the out-of-plane position operator
$\boldsymbol Z=\sum_lz_lP_l\hat{\boldsymbol z}$ and the velocity operator
$\boldsymbol v_s=\hbar^{-1}\boldsymbol\nabla_{\boldsymbol k}h_s^{(0)}$.
The leading-order correction due to the field $\delta h_s^{\rm orb}=-\boldsymbol B_\parallel\cdot
\boldsymbol m_{\parallel,s}^{\rm orb}$ therefore identifies
\begin{equation}
\boldsymbol m_{\parallel,s}^{\rm orb}=-\frac{e}{2}
\left(\boldsymbol Z\times\boldsymbol v_s
-\boldsymbol v_s\times\boldsymbol Z\right),
\label{eq:morb1}
\end{equation}
which is the in-plane orbital magnetic moment \cite{hu2026theory} for the
spin-$s$ sector. First-order
perturbation theory gives
$\Delta_m(\boldsymbol k;B)\approx-\boldsymbol B_\parallel\cdot
\boldsymbol{\mathcal T}_m$, where
$\boldsymbol{\mathcal T}_m=\sum_s s\langle u_{m,s}|
\boldsymbol m_{\parallel,s}^{\rm orb}|u_{m,s}\rangle$ is the susceptibility
vector.

The band sum rule of $\boldsymbol{\mathcal T}_m$ isolates the hidden order
through the layer magnetic dipole
$\boldsymbol F(\boldsymbol k)
=\operatorname{Tr}[\boldsymbol ZM(\boldsymbol k)]$:
\begin{equation}
\sum_m\boldsymbol{\mathcal T}_m(\boldsymbol k)
=
\frac{2e}{\hbar}
\nabla_{\boldsymbol k}\times
\boldsymbol F(\boldsymbol k).
\label{eq:multilayer_tomography_sum_rule}
\end{equation}
Importantly, the sum rule admits a direct representation in terms of the spectral function. The first energy moment of the spin-difference spectral function $A_{\rm sd}$ obeys
\begin{equation}
\mathcal M_{\rm sd}(\boldsymbol k;B)\equiv\int_{-\infty}^{\infty}d\omega\,
\omega A_{\rm sd}(\boldsymbol k,\omega;B)
=
\sum_m\Delta_m(\boldsymbol k;B).
\label{eq:spectral_moment_identity}
\end{equation}
Combining this identity with
Eq.~\eqref{eq:multilayer_tomography_sum_rule} gives, in the weak-field
regime,
\begin{equation}
\mathcal M_{\rm sd}(\boldsymbol k;B)
=
-\frac{2e}{\hbar}
\boldsymbol B_\parallel\cdot
\left[
\nabla_{\boldsymbol k}\times
\boldsymbol F(\boldsymbol k)
\right]
+O(B^3).
\label{eq:spectral_moment_sum_rule}
\end{equation}
For the model in Eq.~\eqref{eq:minimal_M_blocks},
$\boldsymbol F=J\operatorname{Tr}(\boldsymbol Z\Lambda)f_\alpha
=-f_\alpha JdN\hat{\boldsymbol z}$. Thus, measuring the spin-difference spectral moment $\mathcal M_{\rm sd}$ allows reconstruction of the momentum profile of the hidden form factor $f_\alpha(\boldsymbol k)$, without determining every individual band spin splitting.

\begin{figure}[t]
    \centering
    \includegraphics[width=\linewidth]{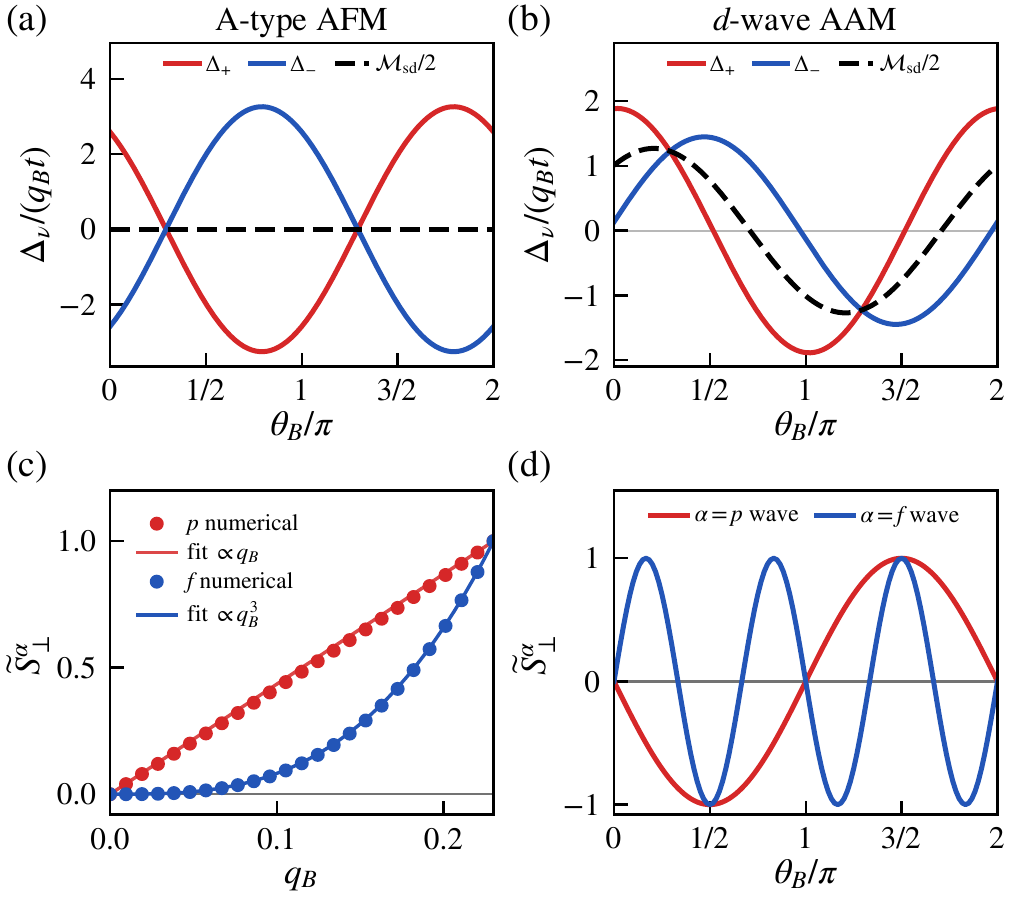}
    \caption{Diagnostics of the order-parameter wave character in a bilayer AAM. (a), (b) The spin-difference spectral moment $\mathcal M_{\rm sd}$ (black dashed) vanishes identically for the A-type AFM in (a) but is generically nonzero for the $d$-wave AAM in (b), distinguishing the two without layer resolution. Red and blue curves show $\Delta_\nu(\theta_B)$ for $\nu=+$ and $\nu=-$, respectively, at $(k_x,k_y)=(0.7,1.0)$. Here $\Delta_\nu$ and $\mathcal M_{\rm sd}$ are evaluated in the $q_B\rightarrow0$ limit. (c), (d) Normalized field-induced out-of-plane spin polarization  $\tilde S_\perp^\alpha$ for odd-parity AAM, with $S_\perp$ normalized by its maximum absolute value. (c) Field-strength dependence at $\theta_B=3\pi/2$: $p$-wave order is linear in $q_B$, whereas $f$-wave order is cubic (circles, exact numerical data; solid lines, fits). (d) Angular dependence at $q_B=0.18$, showing first and third harmonics for $p$- and $f$-wave order, respectively. Panels (c) and (d) use a triangular lattice at $T=0$. Other parameters are the same as in Fig.~\ref{fig:fig1}.}
    \label{fig:diagnostics}
\end{figure}

In particular, in the even-parity sector of a bilayer AAM, the relation in Eq.~\eqref{eq:spectral_moment_sum_rule} yields a null test that distinguishes an
A-type AFM from a $d$-wave AAM. For
$\boldsymbol B_\parallel
=B(\cos\theta_B,\sin\theta_B,0)$, we have
\begin{equation}
\mathcal M_{\rm sd}
\approx
4Jq_B(-\sin\theta_B,\cos\theta_B,0)
\cdot\boldsymbol\nabla_{\boldsymbol k}f_\alpha.
\label{eq:bilayer_spectral_moment}
\end{equation}
This moment therefore vanishes identically for the
momentum-independent A-type AFM, while at a generic fixed momentum it
exhibits a first angular harmonic for the $d$-wave AAM, as shown in
Figs.~\ref{fig:diagnostics}(a) and (b).

\emph{Field-induced spin polarization in odd-parity AAM.}---For the odd-parity sector, a complementary diagnostic is the field-induced net spin polarization. As noted above, the odd-to-even conversion generally permits such a
polarization under an in-plane orbital field:
\begin{equation}
S_\perp(\boldsymbol B_\parallel)
=
\frac12\sum_{\nu,s}s\,
\left\langle
n_F[E_{\nu,s}(\boldsymbol k;\boldsymbol B_\parallel)]
\right\rangle_{\rm BZ},
\end{equation}
in units of $\hbar$ per unit cell. Because $\Delta_m(\boldsymbol k;B)$ is odd in $\boldsymbol B_\parallel$, $S_\perp$ satisfies $S_\perp(-\boldsymbol B_\parallel)=-S_\perp(\boldsymbol B_\parallel)$. The weak-field expansion can therefore contain only odd powers of
$\boldsymbol B_\parallel$. Unlike the splitting itself, $S_\perp$ is a Fermi-surface property and requires a metallic
state (\textbf{End Matter App.~B}).

For the \(p_x\)-wave AAM in the bilayer model of Eq.~\eqref{eq:bilayer_field_hamiltonian}, the order-parameter form factor selects an in-plane direction. So a linear response is symmetry allowed, $S_\perp^{(p)}=\boldsymbol\chi_p\cdot\boldsymbol q_B
+O(q_B^3)$,
giving a first angular harmonic. For the $f$-wave case, by contrast,
both the order-parameter form factor and $S_\perp$ are invariant under $C_{3z}$, requiring
$S_\perp(\boldsymbol q_B)=S_\perp(C_{3z}\boldsymbol q_B)$.
The leading symmetry-allowed term is therefore cubic
\begin{equation}
S_\perp^{(f)}
=
C_fq_B^3
\cos(3\theta_B+\delta_f)
+O(q_B^5),
\end{equation}
and has a third angular harmonic ($2\pi/3$ period with respect to $\theta_B$).

Direct numerical calculations confirm both scalings. We use the bilayer spectrum
of Eq.~\eqref{eq:bilayer_exact_spectrum} on a triangular lattice with
$\varepsilon=-2t[\cos k_x+2\cos(k_x/2)\cos(\sqrt3k_y/2)]-\mu$,
$f_p=\frac23[\sin k_x+\sin(k_x/2)\cos(\sqrt3k_y/2)]$, and
$f_f=\sin k_x-2\sin(k_x/2)\cos(\sqrt3k_y/2)$.
Near $\Gamma$, $f_p\simeq k_x$ whereas
$f_f\simeq-(k_x^3-3k_xk_y^2)/8$.
As shown in Figs.~\ref{fig:diagnostics}(c) and (d), the net spin polarization exhibits the
predicted linear and cubic field dependence with first and third
angular harmonics, respectively. These signatures directly distinguish $p$- and $f$-wave odd-parity AAM orders.

\emph{Conclusion and discussion.}---We have shown that an in-plane orbital magnetic field converts a layer-compensated exchange texture hidden by center exchange into a spin-resolved spectral response. For a parent order of definite momentum parity $\chi$ and an approximately conserved collinear spin, the induced splitting satisfies $\Delta_m(-\boldsymbol k;B)=-\chi\Delta_m(\boldsymbol k;B)$, reversing the parity of the observable response without changing the underlying magnetic order. This exact relation provides the primary readout. Complementary weak-field diagnostics resolve the wave character: we propose a spectral sum rule and demonstrate how to use it to distinguish momentum-independent A-type AFM order from $d$-wave AAM order by employing a minimal model, while the field scaling and angular harmonics of the induced spin polarization distinguish representative $p$- and $f$-wave AAM orders. Together, these observables provide a bulk-sensitive probe of hidden exchange order without local resolution.

The analysis above focuses on the orbital effect and does not include the Zeeman effect. Here we argue that in experiments, the Zeeman coupling will not obscure the orbital-field-induced splitting, and does not invalidate the application of the parity and wave-character diagnostics. First, the two contributions have distinct symmetry signatures: the Zeeman splitting is even in momentum and independent of the in-plane field direction, whereas the orbital response can be identified through its momentum parity or angular dependence. Moreover, a transverse Zeeman coupling leaves both the parity relation in Eq.~\eqref{eq:main_spectral_function} and the spectral moment $\mathcal M_{\rm sd}$ unchanged. Furthermore, the multilayer geometry favors the orbital contribution: its ratio to the projected Zeeman splitting grows as $N^2$ before interlayer coherence limits the enhancement \cite{li2026transdimensional}. For the representative six-layer film, the estimated orbital splitting can exceed the projected Zeeman splitting by nearly two orders of magnitude (see \textbf{End Matter App.~C}).

\emph{Acknowledgments}---The author thanks Xilin Feng, Xun-Jiang Luo, and Jin-Xin Hu for inspiring discussions.

\bibliographystyle{apsrev4-1}
\bibliography{main}

\clearpage
\onecolumngrid
\begin{center}
    \textbf{\large End Matter}\\[.2cm]
\end{center}
\twocolumngrid
\setcounter{equation}{0}
\setcounter{figure}{0}
\setcounter{table}{0}
\renewcommand{\theequation}{S\arabic{equation}}
\renewcommand{\theHequation}{S\arabic{equation}}
\renewcommand{\thefigure}{S\arabic{figure}}
\renewcommand{\thetable}{S\arabic{table}}
\renewcommand{\tablename}{Supplementary Table}

\noindent\emph{Appendix A: Symmetry conditions.} Let $D_{\mathcal S}$ denote the spatial part of $\mathcal S$ in
orbital-layer space. Because its spatial component reverses momentum without exchanging layers, for an arbitrary in-plane momentum $\boldsymbol p$, a unitary
realization $[U_{\mathcal S}\Vert g_{\mathcal S}]$ satisfies
$D_{\mathcal S}h_0(\boldsymbol p)D_{\mathcal S}^\dagger=h_0(-\boldsymbol p)$,
$D_{\mathcal S}M(\boldsymbol p)D_{\mathcal S}^\dagger=\chi M(-\boldsymbol p)$,
and $U_{\mathcal S}\sigma_zU_{\mathcal S}^\dagger=\chi\sigma_z$.
For an antiunitary realization $[U_{\mathcal S}K\Vert g_{\mathcal S}]$, with
$K$ complex conjugation, $h_0$, $M$, and $\sigma_z$ on the left-hand sides
are replaced by their complex conjugates. Setting
$\boldsymbol p=\boldsymbol k+\boldsymbol q_{ll'}(B)$ then gives, in the
unitary case,
\begin{equation}
D_{\mathcal S}h_s^B(\boldsymbol k)
D_{\mathcal S}^\dagger
=h_{\chi s}^{-B}(-\boldsymbol k).
\label{eq:main_S_unitary_field}
\end{equation}
For an antiunitary realization, $h_s^B$ on the left-hand side is replaced
by $(h_s^B)^*$. Thus, $\mathcal S$ reverses the applied field while mapping
the spin block $s$ to $\chi s$. Blockwise, this follows from
$-\boldsymbol p=-\boldsymbol k+\boldsymbol q_{ll'}(-B)$ for every in-plane
field direction. In the full spin space, the two factors of $\chi$, from
the orbital exchange matrix and the spin rotation, leave
$M(-\boldsymbol p)\otimes\sigma_z$ invariant. In the spin-block notation, the same transformation is encoded by $s\to\chi s$, realizing $\mathcal B$ for $\chi=+1$ and $\mathcal C$ for $\chi=-1$. In either realization, an eigenstate $|u\rangle$ is mapped to
$\mathcal G|u\rangle$, with complex conjugation included explicitly in $\mathcal G$, so Eq.~\eqref{eq:main_G_fixed_field} guarantees identical spectra at the symmetry-related momenta.

For the odd-to-even route, the induced splitting is even in momentum, so its
Brillouin-zone sum is not forced to vanish and a net spin polarization
$S_\perp$ is generically allowed. Compensation requires a residual, layer-preserving,
spin-flipping element $[C_2^\perp\Vert g]$ with momentum
action $R_g$ and representation $D_g$ satisfying, at zero field,
\begin{equation}
D_gh_0(\boldsymbol p)D_g^\dagger=h_0(R_g\boldsymbol p),\,
D_gM(\boldsymbol p)D_g^\dagger=-M(R_g\boldsymbol p),
\label{eq:app_Rg_zero_field}
\end{equation}
where the minus sign in the second relation reflects the spin flip. This element remains a symmetry in the
field when the Peierls shifts are left invariant,
$R_g\boldsymbol q_{ll'}=\boldsymbol q_{ll'}$, equivalently
$R_g(\boldsymbol B_\parallel\times\hat{\boldsymbol z})
=\boldsymbol B_\parallel\times\hat{\boldsymbol z}$. Setting
$\boldsymbol p=\boldsymbol k+\boldsymbol q_{ll'}$ then gives
\begin{equation}
D_gh_s^B(\boldsymbol k)D_g^\dagger=h_{-s}^{B}(R_g\boldsymbol k),
\label{eq:app_Rg_field}
\end{equation}
with an explicit complex conjugation for an antiunitary realization, hence
$E_{m,s}(\boldsymbol k;B)=E_{m,-s}(R_g\boldsymbol k;B)$ and
$\Delta_m(R_g\boldsymbol k;B)=-\Delta_m(\boldsymbol k;B)$. The Brillouin zone sum therefore gives $S_\perp=0$.

Because $\boldsymbol B_\parallel$ is an axial vector, a vertical mirror with in-plane
normal $\hat{\boldsymbol n}$ preserves it only for
$\boldsymbol B_\parallel\parallel\hat{\boldsymbol n}$, whereas $C_{2z}$ and
$\mathcal R_z$ reverse it, while the excluded cases $R_g=\pm\mathbb I$ are
$\mathcal A$ and $\mathcal C$ themselves, both already broken by the field.
Counting the opposite-spin mirrors therefore fixes the angular pattern of
$S_\perp$. The $p_x$-wave parent has only $[C_2^\perp\Vert m_x]$ and remains compensated only for
$\boldsymbol B_\parallel\parallel\pm\hat{\boldsymbol x}$, giving two nodes in
$\theta_B$ and a first harmonic. A $C_{3z}$-invariant $f$-wave parent
has three $C_{3z}$-related mirrors, giving six nodes spaced by $\pi/3$ and
a third harmonic. The angular periods of Fig.~\ref{fig:diagnostics}(d) thus
follow from symmetry alone.

\noindent\emph{Appendix B: Weak-field expansion.} For the model in Eq.~\eqref{eq:minimal_h0_blocks}, the susceptibility vector can be obtained analytically as
\begin{equation}
\boldsymbol{\mathcal T}_{n,\nu}^{\alpha}
=
\frac{Jed}{\hbar}
\frac{\tan^2\theta_n}{2N+1}
\hat{\boldsymbol z}\times
\left(
\boldsymbol\nabla_{\boldsymbol k}f_\alpha
+
\frac{\nu f_\alpha}{R_n}
\boldsymbol\nabla_{\boldsymbol k}\varepsilon
\right),
\label{eq:band_resolved_splitting_susceptibility}
\end{equation}
which separates the shape of the spin splitting from its magnitude. The expression in parentheses contains no explicit layer index: for every subband the momentum pattern of the splitting is fixed by the order-parameter form factor $f_\alpha$ and
the intralayer dispersion $\varepsilon$. Within the parentheses, the remaining thickness
dependence enters through $R_n$. The thickness also enters through the
prefactor $\tan^2\theta_n/(2N+1)$, which increases monotonically with $n$. The prefactor is largest for the top subband $n=N$, where it approaches $4(2N+1)/\pi^2$ and therefore scales linearly with the layer number.

The net spin polarization follows from a parallel expansion. Suppressing the band
index, write $E_s=E^{(0)}+u+sw$ with $u=\overline{\delta E}$ even in $B$ and
$w=\Delta/2$ odd. Zero-field spin degeneracy eliminates the zeroth-order term, and
the spin sum becomes
\begin{align}
&\sum_s s\,n_F\!\left(E^{(0)}{+}u{+}sw\right)\nonumber\\
&=n_F'\Delta+n_F''u\Delta+\tfrac{1}{24}n_F'''\Delta^3+O(B^5).
\label{eq:app_spin_sum_telescope}
\end{align}
This expression follows by expanding first in $w$ and then in $u$, with the derivatives of the Fermi function
evaluated at $E^{(0)}$. Sorting by field power with
$\Delta=\Delta^{(1)}+\Delta^{(3)}+\cdots$ and $u=u^{(2)}+\cdots$ gives, up to
$O(B^5)$,
\begin{equation}
\begin{split}
S_\perp&=\tfrac{1}{2}\sum_m\big\langle
n_F'\Delta^{(1)}+n_F'\Delta^{(3)}\\
&+n_F''u^{(2)}\Delta^{(1)}+\tfrac{1}{24}n_F'''(\Delta^{(1)})^{3}\big\rangle_{\rm BZ},
\end{split}
\label{eq:app_Sperp_expansion}
\end{equation}
the first term being $O(B)$ and the remaining three $O(B^3)$.
With $\Delta_m^{(1)}=-\boldsymbol B_\parallel\cdot\boldsymbol{\mathcal T}_m$
the first term is the Fermi-surface average of the same susceptibility vector
that Eq.~\eqref{eq:band_resolved_splitting_susceptibility} gives band by band,
\begin{equation}
S_\perp^{(1)}=-\frac{1}{2}\boldsymbol B_\parallel\cdot\sum_m
\big\langle n_F'(E_m^{(0)})\,\boldsymbol{\mathcal T}_m\big\rangle_{\rm BZ},
\label{eq:app_Sperp_linear}
\end{equation}
or $S_\perp^{(1)}=\boldsymbol\chi\cdot\boldsymbol q_B$ with
$\boldsymbol\chi=-(\hbar/ed)\sum_m\langle n_F'(E_m^{(0)})\,
\boldsymbol{\mathcal T}_m\times\hat{\boldsymbol z}\rangle_{\rm BZ}$. It already carries the parity
rule: for an even-parity parent $\Delta_m$ is odd in $\boldsymbol k$ and
averages to zero over the inversion-symmetric Fermi surface, so only the
odd-to-even route polarizes. The $O(B^3)$ terms are small corrections while the
linear response survives, as for the $p_x$ parent. For a $C_{3z}$-invariant
$f$-wave parent $S_\perp^{(1)}$ vanishes identically, no in-plane vector being
$C_{3z}$ invariant, and the whole $q_B^3$ response comes from
those three terms, none of which may be dropped.

Every weight in Eq.~\eqref{eq:app_Sperp_expansion} is a derivative of $n_F$,
so the response is a Fermi-surface property at every order. For a gapped AAM
the corresponding statement is nonperturbative: all occupied bands are filled, the two spin
sectors contribute equally, and $S_\perp$ vanishes for any field that does not
close the gap, whereas $\Delta_m(\boldsymbol k)$ exists throughout the Brillouin zone
irrespective of filling.

\noindent\emph{Appendix C: Zeeman effect and interlayer coherence.} We finally
examine the Zeeman effect of the multilayer model and
compare it with the orbital-field-induced splitting discussed in the
\textbf{Main Text}. The following discussion assumes that the conserved collinear spin axis lies along \(z\) and that an in-plane magnetic field produces the Zeeman term $\mathcal H_Z=-(g_s\mu_B/2)\boldsymbol B_\parallel\cdot\boldsymbol{\sigma}_\parallel$.

For the $(n,\nu)$ branch of the model in
Eqs.~\eqref{eq:minimal_h0_blocks} and \eqref{eq:minimal_M_blocks},
the weak-field Zeeman splitting is
\begin{equation}
\Delta^{Z}_{n,\nu}(\boldsymbol k)
=
\mu_B B_\parallel g_s
\frac{2t_\perp|\cos\theta_n|}{R_n}
+O(B_\parallel^3),
\label{eq:zeeman_splitting}
\end{equation}
which is independent of the direction of $\boldsymbol B_\parallel$ and
even in $\boldsymbol k$. Thus it can be distinguished from the orbital
effect by its field-angle dependence in odd-to-even
routes and by its momentum parity for even-to-odd routes,
respectively. The factor
$2t_\perp|\cos\theta_n|/R_n$ reduces the effective $g$ factor through the
hidden spin texture, most strongly at $n=N$, where
$2t_\perp\cos\theta_N\simeq\pi t_\perp/(2N+1)$ at large $N$.
Away from nodes of the order-parameter form factor, where
$|Jf_\alpha(\boldsymbol k)|\gg2t_\perp|\cos\theta_N|$, this gives
$\Delta^Z_{N,\nu}\propto B_\parallel/N$. Since $|\Delta^{\rm orb}_{N,\nu}|\propto B_\parallel N$ from
Eq.~\eqref{eq:band_resolved_splitting_susceptibility}, the ratio
$|\Delta^{\rm orb}_{N,\nu}|/|\Delta^{Z}_{N,\nu}|$ grows as $N^2$.

This enhancement cannot persist to arbitrarily large $N$, since the
layer-hybridized response requires coherence across the film with the coherence length $\xi_z\simeq2t_\perp d\tau/\hbar$~\cite{mckenzie1998incoherent,wang2004quasiparticle,li2026transdimensional,hu2026theory}, where $\tau$ is the quasiparticle lifetime. For a $2N$-layer film, this yields $N\lesssim N_{\rm coh}
\simeq t_\perp\tau/\hbar$. Interlayer coherence thus cuts off the large-$N$ enhancement.

For an order-of-magnitude estimate, we take the parameters from the recently reported AAM candidate $\mathrm{Cs}_{1-\delta}\mathrm{V}_2\mathrm{Te}_2\mathrm{O}$ \cite{yang2025observation}.
Single-crystal X-ray diffraction gives an in-plane lattice constant
$a=4.043\,\mathring{\rm A}$ and an out-of-plane lattice constant
$c=8.853\,\mathring{\rm A}$, while the adjacent altermagnetic sectors are
separated by $d\simeq0.89\,{\rm nm}$ and exhibit only weak interlayer
hybridization~\cite{CsStructure,yang2025observation}. A recent low-energy description of
$\mathrm{Cs}\mathrm{V}_2\mathrm{Te}_2\mathrm{O}$ uses an
interlayer hopping $t_\perp\simeq10\,{\rm meV}$~\cite{CsModel}. We retain a
quasiparticle scattering rate $\hbar/\tau=2\,{\rm meV}$ as a representative
clean-film value, which gives
$N_{\rm coh}\simeq t_\perp\tau/\hbar\sim5$. We therefore use a six-layer film
($N=3$) as a conservative benchmark. Restoring the lattice constant,
the dimensionless Peierls shift is
$q_Ba=eB_\parallel da/(2\hbar)$, which gives
$q_Ba=2.72\times10^{-3}$ at $B_\parallel=10\,{\rm T}$. Retaining
$J=0.2\,{\rm eV}$ as the effective $d$-wave exchange amplitude of our minimal
model, we obtain
$|\Delta^{\rm orb}_{3,\nu}|\sim3.0\,{\rm meV}$. At a generic
momentum away from the order-parameter nodes, with
$|f_\alpha(\boldsymbol{k})|\sim1$ and $g_s=2$, the bare Zeeman scale
$g_s\mu_BB_\parallel\simeq1.16\,{\rm meV}$ is strongly suppressed by the reduced effective transverse \(g\) factor of the relevant band, yielding
$|\Delta^Z_{3,\nu}|\simeq0.026\,{\rm meV}$. The corresponding
ratio, $|\Delta^{\rm orb}_{n,\nu}|/|\Delta^Z_{n,\nu}|\sim1.2\times10^2$, shows that the orbital
contribution can exceed the projected Zeeman splitting by more than two
orders of magnitude over the experimentally relevant few-layer regime. 

For the spin-difference spectral moment, Eq.~\eqref{eq:spectral_moment_sum_rule}, together with
$\boldsymbol F=-JdNf_\alpha\hat{\boldsymbol z}$, gives
\begin{equation}
\mathcal M_{\rm sd}
\simeq
4JNq_B
(-\sin\theta_B,\cos\theta_B,0)
\cdot\nabla_{\boldsymbol k}f_\alpha.
\label{eq:Mr_estimate}
\end{equation}
With the parameters above, $q_Ba=2.72\times10^{-4}$
already at $B_\parallel=1$ T. For the $d$-wave form factor $f_d=\cos k_x-\cos k_y$, we obtain $|\mathcal M_{\rm sd}|_{\rm max}\simeq4JN(q_Ba)\sqrt{2}$. The resulting spectral moment is therefore
$|\mathcal M_{\rm sd}|\simeq0.92$ meV for a six-layer film ($N=3$) and $\simeq1.54$ meV for a ten-layer film ($N=5$). Thus, within the coherent few-layer regime, a Tesla-scale in-plane field can generate a spin-difference spectral moment of order $1$~meV.

\end{document}